\documentclass[lettersize,journal]{IEEEtran}
\usepackage{amsmath,amsfonts}
\usepackage{algorithmic}
\usepackage{array}
\usepackage[caption=false,font=small,labelfont=rm,textfont=rm]{subfig}
\usepackage{textcomp}
\usepackage{stfloats}
\usepackage{url}
\usepackage{verbatim}
\usepackage{graphicx}
\def\BibTeX{{\rm B\kern-.05em{\sc i\kern-.025em b}\kern-.08em
    T\kern-.1667em\lower.7ex\hbox{E}\kern-.125emX}}
\usepackage{balance}
\usepackage[utf8]{inputenc}
\usepackage[ruled,vlined]{algorithm2e}
\usepackage{xcolor}
\usepackage[switch]{lineno}
\begin{document}
\title{Semantic-Aware Resource Allocation Based on Deep Reinforcement Learning for 5G-V2X HetNets}
\author{Zhiyu Shao, Qiong Wu,~\IEEEmembership{Senior Member,~IEEE}, Pingyi Fan,~\IEEEmembership{Senior Member,~IEEE}, \\Nan Cheng,~\IEEEmembership{Senior Member,~IEEE}, Qiang Fan, and Jiangzhou Wang,~\IEEEmembership{Fellow,~IEEE}
\thanks{This work was supported in part by the National Natural Science Foundation of China under Grant No. 61701197, in part by the National Key Research and Development Program of China under Grant No.2021YFA1000500(4), in part by the 111 Project under Grant No. B12018. (Corresponding authors: Qiong Wu; Pingyi Fan.)
	
	Zhiyu Shao and Qiong Wu are with the School of Internet of Things Engineering, Jiangnan University, Wuxi 214122, China (e-mail: zhiyushao@stu.jiangnan.edu.cn, qiongwu@jiangnan.edu.cn)
	
	Pingyi Fan is with the Department of Electronic Engineering, Beijing National Research Center for Information Science and Technology, Tsinghua University, Beijing 100084, China (e-mail: fpy@tsinghua.edu.cn).
	
	Nan Cheng is with the State Key Lab. of ISN and School of Telecommunications Engineering, Xidian University, Xi’an 710071, China (e-mail: dr.nan.cheng@ieee.org).
	
	Qiang Fan is with Qualcomm, San Jose, CA 95110, USA (e-mail: qf9898@gmail.com).
	
	Jiangzhou Wang is with the School of Engineering, University of Kent,
	CT2 7NT Canterbury, U.K. (email: j.z.wang@kent.ac.uk).}}

%\markboth{Journal of \LaTeX\ Class Files,~Vol.~18, No.~9, September~2020}%
%{How to Use the IEEEtran \LaTeX \ Templates}

\maketitle

\begin{abstract}
This letter proposes a semantic-aware resource allocation (SARA) framework with flexible duty cycle (DC) coexistence mechanism (SARADC) for 5G-V2X Heterogeneous Network (HetNets) based on deep reinforcement learning (DRL) proximal policy optimization (PPO).  
Specifically, we investigate V2X networks within a three-tiered HetNets structure. 
To meet the demands of high-speed vehicular networking in urban environments, we design a semantic communication system and introduce two resource allocation metrics: high-speed semantic transmission rate (HSR) and semantic spectrum efficiency (HSSE).
Additionally, we address the coexistence of vehicular users and WiFi users in 5G New Radio Unlicensed (NR-U) networks. 
Our approach jointly optimizes the DC coexistence mechanism and the allocation of resources and base stations (BSs).  
Unlike traditional bit-based transmission methods, our approach integrates the semantic communication into the communication system. 
Experimental results show that our proposed framework significantly improves HSSE and semantic throughput (ST) for both vehicular and WiFi users compared to conventional methods. 
\end{abstract}

\begin{IEEEkeywords}
Semantic communication, vehicular networks, resource allocation, unlicensed spectrum bands, 5G NR-U coexistence, deep reinforcement learning.
\end{IEEEkeywords}

\section{Introduction}
\IEEEPARstart{T}{he} fifth-generation (5G) networks aim to provide high-speed, low-latency, and reliable communication services for applications like vehicle-to-everything (V2X) communications \cite{r1}. However, increasing devices pose challenges in capacity and spectrum efficiency. In urban areas, traditional networking mechanisms may fall short. Deploying small cells and heterogeneous networks (HetNets) enhances network capacity by increasing antennas on smaller base stations (BSs)\cite{r2}.
% \cite{RN102}

Recent advancements in semantic communication focus on the meaning of information rather than raw data, improving spectrum efficiency \cite{r3}. Studies have explored its application in various domains, including images, text, audio, mixed reality services, and wireless sensing \cite{r4, r5, r6, r12, r13}.

5G New Radio Unlicensed (NR-U) technology expands 5G HetNets' capacity by operating in unlicensed bands \cite{r7}. Coexistence mechanisms with other wireless networks like WiFi, such as listen before talk (LBT) \cite{r8} and carrier sensing adaptive transmission (CSAT) \cite{r9}, must be considered due to interference.
However, the uncertainty of channel access in NR-U leads to worse performance compared to licensed bands, and thus traditional NR-U and WiFi coexistence mechanisms pose a new challenge for 5G networks.

Furthermore, traditional bit-based resource allocation in vehicular networks ignores semantic information, causing poor spectrum utilization and congestion \cite{r14}. Although deep reinforcement learning (DRL) offers dynamic optimization \cite{r15}, it still relies on a bit-based framework. Improving spectrum efficiency requires a semantic approach and addressing NR-U and WiFi coexistence issues. While \cite{r10} addresses semantic communication resource allocation, it lacks a comprehensive semantic-aware framework for 5G-V2X HetNets and does not adapt well to the dynamic nature of vehicular networks or address the joint optimization of duty cycles and resource allocation.
 
To handle the above issue, this letter introduces semantic communication into a three-tier HetNet vehicular communication system in urban environments for the first time. 
We propose a flexible duty cycle (DC) approach to address coexistence issues between vehicular and WiFi users in NR-U networks, introducing high-speed semantic transmission rate (HSR) and semantic spectrum efficiency (HSSE) metrics in the resource allocation optimization. 
We design a semantic-aware resource
allocation framework with flexible DC
coexistence mechanism (SARADC) algorithm, applying proximal policy optimization (PPO) DRL to optimize flexible DC and resource allocation based on semantic awareness to maximize semantic throughput (ST) \footnote{The source code has been released at: https://github.com/qiongwu86/Semantic-Aware-Resource-Allocation-Based-on-Deep-Reinforcement-Learning-for-5G-V2X-HetNets}.
Experimental results demonstrate that our proposed algorithm outperforms other baselines in terms of HSSE and ST for both vehicular and WiFi users.   

\section{System Model}
\begin{figure}[htbp]
	\centering
	\vspace{-0.35cm}
	\includegraphics[width=0.75\columnwidth]{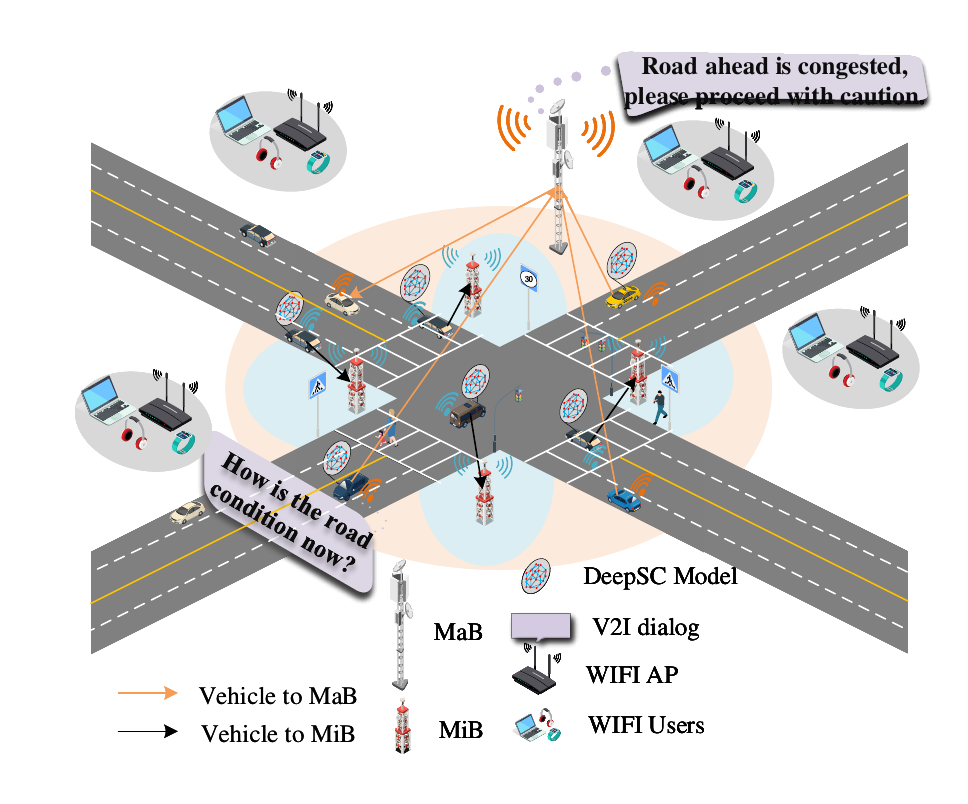}
	\caption{The semantic-aware resource allocation system}
	\vspace{-0.35cm}
	\label{fig1}
\end{figure}

As shown in Fig.\ref{fig1}, we focus on V2X communication in a high-speed urban environment with $N$ vehicles moving in varying directions at a constant speed $V$, resulting in $N$ V2I links.  
The channel state updates every ${t_u}$ ms.
We consider a three-tier HetNets deployment: tier 1 comprises ${B_1}$ macro base stations (MaBs), providing broad coverage and high-capacity communication in licensed frequency bands; 
tier 2 includes ${B_2}$ micro base stations (MiBs) operating in unlicensed frequency bands, offering high-speed and low-latency services;
tier 3 incorporates ${\rm P}$ WiFi access points (APs), also utilizes unlicensed bands. 
${\rm B} = \left\{ {B_1, B_2} \right\}$ represents BSs, with MaBs and MiBs each having ${R_1}$ and ${R_2}$ resource blocks (RB), respectively. 
The $N$ vehicles, ${\rm B}$ BSs and ${\rm P}$ WiFi APs are equipped with DeepSC models \cite{r10}.
DeepSC model includes a transmitter, channel and receiver. The transmitter includes semantic and channel encoding, while the receiver encompasses semantic and channel decoding using the transformer model with shared background knowledge.
For example, a DeepSC-equipped vehicle encodes "How is the road condition?" into semantic information, which MaB decodes back to "Road ahead is congested, please proceed with caution."
The model is pre-trained on MaB, and then the trained semantic DeepSC transmitter models are broadcast and employed by all vehicles due to the complex semantic information extraction during training process.
\vspace{-0.15cm}
\subsection{DeepSC Transceivers and Novel Metrics}
The $n$-th transmitter vehicle generates a sentence ${S_n}$ with ${l_q}$ words, where 
${S_n} = \left[ {{s_{n,1}},{s_{n,2}}, \ldots ,{s_{n,l}}, \ldots ,{s_{n,{l_q}}}} \right]$.
This sentence is processed by the DeepSC encoding part to extract semantic information ${X_n}$ from ${S_n}$
\begin{equation}
	\label{eq1}
	{X_n} = c{h_\beta }\left( {s{e_\alpha }\left( {{S_n}} \right)} \right),
\end{equation}
where $s{e_\alpha }{\rm{ }}\left( \cdot \right)$ and $c{h_\beta }\left( \cdot \right)$ are the semantic and channel encoder networks with parameters $\alpha$ and $\beta$ , respectively.
The semantic symbol vector is ${X_n} = \left[ {{x_{n,1}},{x_{n,2}}, \ldots ,{x_{n,u{l_q}}}} \right]$, where ${u}$ represents the average number of semantic symbols used for each word.
Then the encoded semantic information is transmitted via the wireless channel to get the received semantic signal
\begin{equation}\label{eq2}
	{Y_n} = {H_{n,i}}{X_n} + {\varsigma ,}
\end{equation}
where ${H_{n,i}}$ is the channel gain for the $i$-th V2I link, $\varsigma$ is the noise.
So the signal-to-interference-noise Ratio (SINR) for the $n$-th vehicle is
\begin{equation}\label{eq3}
	SIN{R_{n,b,r}} = \frac{{{\eta _{n,b,r}}{P_{n,b,r}}{H_{n,b,r}}}}{{{D_{n,b,r}} + {\delta ^2}}},
\end{equation}
where ${\eta _{n,b,r}}$ is a binary indicators, and ${\eta _{n,b,r}} = 1$ means that $n$-th vehicle transmits semantic data to $b$-th BS using $r$-th RB.
${P_{n,b,r}}$ and ${H_{n,b,r}}$ are the transmit power and channel gain from $n$-th vehicle to $b$-th BS on the $r$-th RB, ${\delta ^2}$ is the noise power.
${D_{n,b,r}}$ is the interference on the $r$-th RB of $b$-th BS incurred by other BS including both MaBs and MiBs, defined as
\begin{equation}\label{eq4}
	{D_{n,b,r}} = \left\{ \begin{array}{l}
		\sum\limits_{\scriptstyle\hat n = 1\hfill\atop
			\scriptstyle\hat n \ne n\hfill}^N {\sum\limits_{\hat b = 1}^{{B_2}} {{\eta _{\hat n,\hat b,r}}{P_{\hat n,\hat b,r}}{H_{\hat n,b,r}}{\rm{   }}\quad if\hspace{0.1cm}{\rm{  }}b \in {B_1}} } \\
		\sum\limits_{\scriptstyle\hat n = 1\hfill\atop
			\scriptstyle\hat n \ne n\hfill}^N {\sum\limits_{\hat b = 1}^{{B_1}} {{\eta _{\hat n,\hat b,r}}{P_{\hat n,\hat b,r}}{H_{\hat n,b,r}}{\rm{   }}\quad if\hspace{0.1cm}{\rm{  }}b \in {B_2}} } , 
	\end{array} \right.
\end{equation}
So the decoded signal can be represented as 
\begin{equation}\label{eq5}
	{{\hat S}_n} = se_\mu ^{ - 1}\left( {ch_\nu ^{ - 1}\left( {{Y_n}} \right)} \right),
\end{equation}
where $se_\mu ^{ - 1}{\rm{ }}\left( \cdot \right)$ and $ch_\nu ^{ - 1}\left( \cdot \right)$ are semantic and channel decoder networks with parameters $\mu$ and $\nu$.
To optimize the DeepSC model, it uses Cross-entropy (CE) as one of the loss functions to quantify the difference between the reconstructed ${{\hat S}_n}$ and the original sentence ${S_n}$. Additionally, negative mutual information between the transmitted symbols ${S_n}$ and the received symbols ${Y_n}$ serves as another loss function.
To evaluate the semantic communication performance, we employ semantic similarity as a performance metric $\xi$ using bidirectional encoder representations of transformers (BERT) $B\left( \cdot \right)$ \cite{r11}, given by
\begin{equation}\label{eq6}
	\xi  = \frac{{B({S_n})B({{\hat S}_n})}}{{\left\| {B({S_n})} \right\|\left\| {B({S_n})} \right\|}},
\end{equation}
where $\xi$ indicates the highest similarity between the two sentences ($0\le\xi\le 1$).
Assuming text dataset is $\Upsilon  = \sum\limits_{i = {\rm{1}}}^Z {{S_i}}$, where $Z$ is the total number of sentences. The average semantic information for each sentence is $I = \sum\limits_{i = 1}^Z {{I_i}p\left( {{S_i}} \right)}$, where ${I_i}$ is the semantic information of sentence ${S_i}$. 
$p\left( {{S_i}} \right)$ denotes the probability of sentence ${S_i}$ appearing in the datasets. 
Similarly, the average sentence length is $L = \sum\limits_{i = 1}^Z {{l_i}p\left( {{S_i}} \right)} $. The HSR is defined as
\begin{equation}\label{eq7}
	HSR = W\frac{I}{{{u}L}}\xi,
\end{equation}
where the unit of ${u}$ is semantic unit (sut) \cite{r10},  thus the unit of HSR is $suts/s$, which represents semantic transmission rate, so HSSE is defined as
\begin{equation}\label{eq8}
	HSS{E} = \frac{{HSR}}{W} = \frac{I}{{{u_q}L}}{\xi} ,
\end{equation}
the unit of HSSE is suts/s/Hz, which stands for the efficiency of transmitting semantic information. 
In Eq.~(\ref{eq8}), the value of $\frac{I}{L}$, influenced by the source type, remains constant and is negligible during optimization. 
The parameter $\xi$ is represented as $\xi = \Psi({u}, \text{SINR})$ \cite{r10}.
\vspace{-0.4cm}
\subsection{Flexible DC Mechanism}
\begin{figure}[htbp]
	\centering
	\vspace{-0.1cm}
	\includegraphics[width=0.7\columnwidth]{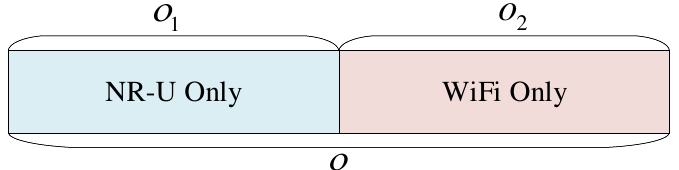}
	\caption{SADC Mechanism}
	\label{fig2}
\end{figure}

Since vehicles use NR-U for semantic transmission and WiFi employs unlicensed spectrum, collisions may occur due to overlapping frequency bands. 
Fig. \ref{fig2} shows that a fixed time slot $o$ is divided into two segments: in ${o_1}$, vehicles exclusively transmit semantic data packets over NR-U, WiFi users access the unlicensed spectrum in ${o_2}$. 
The time durations of ${o_1}$ and ${o_2}$ are adaptable based on network requirements.
According to \eqref{eq7}, the ST of vehicle users is $S{T_n} = HS{R_n} \times {o_1}$. 
As the transmission rate of WiFi users within proximity of $n$-th vehicle is ${R_w}$. The ST of WiFi users according to IEEE 802.11ax (WiFi 6) \cite{r10} is $S{T_w} = \frac{{{R_w}}}{\mu } \times {o_2}$, where $\frac{{{R_w}}}{\mu }$ is $HS{R_w}$ and ${\mu }$ is the number of average semantic symbol per word.
% \cite{RN105}

\subsection{Optimization Problem}
The objective is to maximize the average HSSE by optimizing channel allocation $\eta _{b,r}$, power allocation $p$, the time period ${o_1}$ for vehicles exclusively transmitting semantic data packets over NR-U (Since total time slot $o$ is fixed, optimizing ${o_1}$ automatically determines ${o_2}$) and the number of average semantic symbols per word $\mu$. 
So the problem formulation is 
\begin{subequations}\label{P0}
	\begin{align}
		&{P_0}: \mathop {\max }\limits_{\eta _{b,r} ,p,{o_1},\mu } HSS{E_n}/N \label{eq:11a}\\
		&s.t.\quad\sum\limits_{n = 1}^N {S{T_n}}  \ge \underline {S{T_n}} , \label{eq11b}\\
		&\sum\limits_{n = 1}^N {S{T_w}}  \ge \underline {S{T_w}} , \label{eq11c}\\
		&\sum\limits_{n = 1}^N {{\eta _{n,b,r}} \le 1} \qquad\forall b \in {\rm B},\forall r \in R, \label{eq11d}\\
		&\xi  \ge {\xi _{th}}, \label{eq11e}\\
		&u \in \left\{ {0,1, \cdots ,{u_{\max }}} \right\}. \label{eq11f}
	\end{align}
\end{subequations} 
P0 is fundamentally an optimization problem under specific constraints, aimed at maximizing the average HSSE by determining optimal parameters $\eta _{b,r}$, $p$, $o_1$ and $\mu$.
Constraints (\ref{eq11b}) and (\ref{eq11c}) ensure that both vehicle users and WiFi users meet their minimum thresholds  $\underline {S{T_n}}$ and $\underline {S{T_v}}$, respectively.
Constraint (\ref{eq11d}) ensures that each RB is allocated to at most one vehicle, following OFDMA principles. 
Constraint (\ref{eq11e}) imposes a minimum requirement on semantic similarity ${\xi _{th}}$.
Constraint (\ref{eq11f}) limits the average number of semantic symbols per word within ${{u_{\max }}}$ range.

Due to the non-convex nature of the objective function $HSSE$ in Eq. (\ref{eq8}), dependent on both $\mu$ and $SINR$, modeled as $\xi = \Psi \left( {{u},SINR} \right)$.
Traditional convex optimization methods may converge to suboptimal solutions. Thus, we choose DRL for its ability to handle high-dimensional, nonlinear problems through interactive learning. Specifically, we chose PPO in DRL for effectively tackling non-convex optimization challenges through iterative policy adaptation, well-suited for dynamic vehicular networks.
\begin{algorithm}
	\caption{SARADC Algorithm solving ${P_0}$}
	\label{al1}
	\KwIn{$\pi ,\theta ,{\theta ^v},{\theta _{old}},\theta _{old}^v$ }
	\KwOut{the optimal policy parameter ${\pi ^ * }$}
	
	Initialize parameters ${\theta _{old}} = \theta , \theta _{old}^v = {\theta ^v}$;
	
	Initialize replay experience buffer $R$;
	
	\For{episode from $1$ to ${E_{\max }}$}
	{
		\For{$t$ from $1$ to ${t_{\max }}$}
		{
			Receive observation state $s_n^t$;
			
			Select action $a_n^t = \pi \left( {a_n^t\left| {s_n^t;{\theta _{old}}} \right.} \right)$ using the old policy;
			
			Observe reward $r_n^t$ and next state $s_n^{t + 1}$;
			
			Store tuple $\left( {s_n^t,a_n^t,r_n^t,s_n^{t + 1}} \right)$ in $R$;
			
			Compute advantages ${A_t}$ according to (\ref{eq17})
			
			Normalize ${A_t}$
			
			\If{$t$ can be exactly divided by $t_u$}{
				
				\For{epochs from $1$ to ${K_{\max }}$}
				{
					Randomly sample a batch of $bs$ experience from the replay buffer $R$;
					
					Compute actor loss  according to (\ref{eq18})
					
					Compute critic loss  according to (\ref{eq19})
					
					Compute total loss according to (\ref{eq20})
					
					Update actor-critic parameters $\theta$ and ${\theta ^v}$ according to (\ref{eq21}) and (\ref{eq22});
					
					Update old actor-critic parameters ${\theta _{old}}$ and $\theta _{old}^v$ with $\theta$ and ${\theta ^v}$.
			}}
		}
	}
\end{algorithm}
\section{Proposed SARADC Algorithm Approach}
To address the above problem, we propose a SARADC algorithm that utilize DRL, specifically PPO to optimize resource allocation in high-speed vehicular networks. 
%This approach effectively adapts to dynamic environments and complex state spaces, enhancing communication efficiency through semantic data transmission.
At each time step $t$, the state of each vehicle agent includes parameters such as instant channel gain $h_{n,b,r}^t$, SINR $SINR_{n,b,r}^t$ when connected to $b$-th BS on $r$-th RB, HSSE of the vehicle $HSS{E_n}_{n,b,r}^t$, the HSSE of the WiFi at the previous time $HSS{E_w}_{n,b,r}^{t - 1}$ and the previous interference $D_{n,b,r}^{t - 1}$ from other vehicles to $b$-th BS on $r$-th RB. Thus, the state of each agent can be represented as
\begin{equation}\label{eq12}
	s_n^t = \left[ {h_{n,b,r}^t,SINR_{n,b,r}^t,HSS{E_n}_{n,b,r}^t,HSS{E_w}_{n,b,r}^{t - 1},D_{n,b,r}^{t - 1}} \right] .
\end{equation}
After observing the environment state $s_n^t$, each agent selects an action $a_n^t$ based on policy $\pi$, which includes BS and RB allocation ${\eta _{n,b,r}^t}$, allocated transmission power $p_n^t$, the proportion of DC for vehicle connected to MiBs ${o_1}_n^t$ and the number of semantic symbols represented by each word $\mu _n^t$. Thus, the action of each agent can be represented as
\begin{equation}\label{eq13}
	a_n^t = \left[ {{\rm{ }\eta _{n,b,r}^t},p_n^t,{o_1}_n^t,\mu _n^t} \right].
\end{equation}
After taking action $a_n^t$, each agent receives a reward $r_n^{t}$ to evaluate its behavior, and the environment transits to the next state $s_n^{t + 1}$.
The reward for each agent can be expressed as
\begin{equation}\label{eq14}
	r_n^t = \frac{{\sum\limits_{n = 1}^N {S{T_n}} }}{{N\underline {S{T_n}} }}\omega \left( {S{T_w},\underline {S{T_w}} } \right) - C\psi \left( {\eta _{n,b,r}^t} \right),
\end{equation}
where $\omega$ and $\psi$ are functions ensuring satisfaction of specific constraints satisfies constraint $\omega \left( {x,y} \right) = \left\{ {\begin{array}{*{20}{l}}
		{1,x \ge y}\\
		{0,x < y}
\end{array}} \right.$ and $\psi \left( {\eta _{n,b,r}^t} \right) = \left\{ {\begin{array}{*{20}{l}}
{1,\sum\limits_{r = 1}^R {\eta _{n,b,r}^t} > 1}\\
{0,otherwise}
\end{array}} \right.$.
$C$ is a penalty factor.
Meanwhile, cumulative discounted reward is $R_n^t = \sum\nolimits_{\tau  = 0}^\infty  {{\gamma ^\tau }r_n^{t + \tau }}$, where $\gamma$ $\left( {0 < \gamma  < 1} \right)$is discount factor.

The SARADC algorithm employs PPO, consisting of actor and critic networks. 
Initially, parameters of actor network $\theta$ and critic network ${\theta^v}$ are initialized using a random normal distribution ($\mathcal{N}(0, 0.1)$).
The old parameters ${\theta_{old}}$ and ${\theta_{old}^v}$, are set to the initial values of $\theta$ and ${\theta^v}$, respectively. An experience replay buffer $R$ is also initialized.

The main training loop consists of an outer loop for episodes (${E_{\max}}$), representing the total number of training episodes, and an inner loop running for ${t_{\max}}$, representing the time steps within each episode. 
At each time step $t$, the current state $s_n^t$ is observed, and an action $a_n^t$ is selected based on the old policy network using $a_n^t = \pi \left( {a_n^t\left| {s_n^t;{\theta _{old}}} \right.} \right)$. After executing the action, the reward $r_n^t$ and the next state $s_n^{t+1}$ are observed. The experience tuple $\left( {s_n^t,a_n^t,r_n^t,s_n^{t+1}} \right)$ is then stored in $R$.
Next, the advantage function is computed and normalized based on the cumulative discounted reward $R_n^t$ and the value function estimation ${V\left( {s_n^t;\theta _{old}^v} \right)}$
\begin{equation}\label{eq17}
	{A_t} = \left( {R_n^t - V\left( {s_n^t;\theta _{old}^v} \right)} \right)/{A_{t\max }}.
\end{equation}
If the current time step $t$ is a multiple of the update interval ${t_u}$, the algorithm updates the actor and critic networks through an inner loop from 1 to ${K_{\max}}$. During each epoch, a batch of $bs$ experience is sampled from $R$.
The loss of the actor network is 
\begin{equation}\label{eq18}
	{L^{actor}}(\theta ) =  - \min \left( {{r_t}(\theta ){A_t}, clip \left( {{r_t}(\theta ),1 - \varepsilon ,1 + \varepsilon } \right){A_t}} \right) ,
\end{equation}
where ${r_t}(\theta ) = \exp \left( {\pi \left( {a_n^t\left| {s_n^t;\theta } \right.} \right) - \pi \left( {a_n^t\left| {s_n^t;{\theta _{old}}} \right.} \right)} \right)$ is the ratio of the log probability of new actions to old ones, , $clip\left(  \cdot  \right)$ limits the range of ${{r_t}(\theta ){A_t}}$ by clipping, and $\varepsilon$ is a hyper-parameter that adjusts the clipping fraction.
The loss of the critic network is
\begin{equation}\label{eq19}
	{L^{critic}}({\theta ^v}) = 0.5{\left( {R_n^t - V\left( {s_n^t;{\theta ^v}} \right)} \right)^2}.
\end{equation}
The total loss of the actor-critic network is 
\begin{equation}\label{eq20}
	{L^{total}}(\theta ,{\theta ^v}) = {L^{actor}}(\theta ) + {L^{critic}}({\theta ^v}) - c \times \left( {entropy} \right),
\end{equation}
where $entropy =  - \pi \left( {a_n^t\left| {s_n^t;\theta } \right.} \right) \times \log \left( {\pi \left( {a_n^t\left| {s_n^t;\theta } \right.} \right)} \right)$ is the entropy bonus, and $c$ is the regularization coefficient.
The parameters $\theta$ and ${\theta ^v}$ of the actor and critic networks are updated using gradient descent by minimizing (\ref{eq20})
\begin{equation}\label{eq21}
	\theta  = \theta  - lr \cdot \nabla \theta  \cdot {L^{total}}(\theta ,{\theta ^v})
\end{equation}
\begin{equation}\label{eq22}
	{\theta ^v} = {\theta ^v} - lr \cdot \nabla {\theta ^v} \cdot {L^{total}}(\theta ,{\theta ^v})
\end{equation}
The old parameters ${{\theta _{old}}}$ and ${\theta _{old}^v}$ are updated to match the current values of $\theta$ and ${\theta^v}$. The policy parameter ${\pi}$ is updated directly through the optimization of $\theta$ and indirectly through ${\theta^v}$. 
As these parameters are updated to minimize the total loss function (\ref{eq20}), the optimal policy ${\pi^*}$ is obtained as ${\pi}$ is gradually refined to maximize the expected return.
The computational complexity of the fully connected actor and critic networks is $O\left( {\sum\nolimits_{f = 2}^{F - 1} {{\Gamma _{f - 1}}{\Gamma _f} + } {\Gamma _f}{\Gamma _{f + 1}}} \right)$, where ${{\Gamma _f}}$ represents the number of neural units in the $f$-th layer. 
The input and output layers have units corresponding to the dimensions of the state and action spaces, respectively.
The training process for SARADC scheme is summarized in Algorithm 1.

\section{Simulation Results}
\begin{table}[t]
	\caption{ Parameters of System Model and SARADC}
	\label{tab1}
	\begin{center}
		\begin{tabular}{|c|c|c|c|}
			\hline
			\multicolumn{4}{|c|}{Parameters of System Model}\\
			\hline
			\textbf{Parameter}&{\textbf{Value}}&{\textbf{Parameter}}&{\textbf{Value}} \\
			\hline
			$W$ & 15 KHz & RB number & 12 \\
			\hline
			$u_{\max}$ & 5-25 bits/word & Vehicle height & 1.5 m \\
			\hline
			BS height & 25 m & ${\xi _{th}}$ & $0.9$ \\
			\hline
			
			\hline
			\multicolumn{4}{|c|}{Parameters of SARADC} \\
			\hline
			\textbf{Parameter} & \textbf{Value} & \textbf{Parameter} &  \textbf{Value} \\
			\hline
			$lr$ & $0.0001$ & $\partial$ & $(0.9,0.99)$ \\
			\hline
			$E_{\max}$ & 1000 & $t_{\max}$ & 100 \\
			\hline
			$t_u$ & 5 & $K_{\max}$ & 5 \\
			\hline
			$\gamma$ & 0.99 & $\varepsilon$ & 0.2 \\
			\hline
		\end{tabular}
	\end{center}
\end{table}
\begin{figure}[t]
	\centering
	\includegraphics[width=0.75\columnwidth]{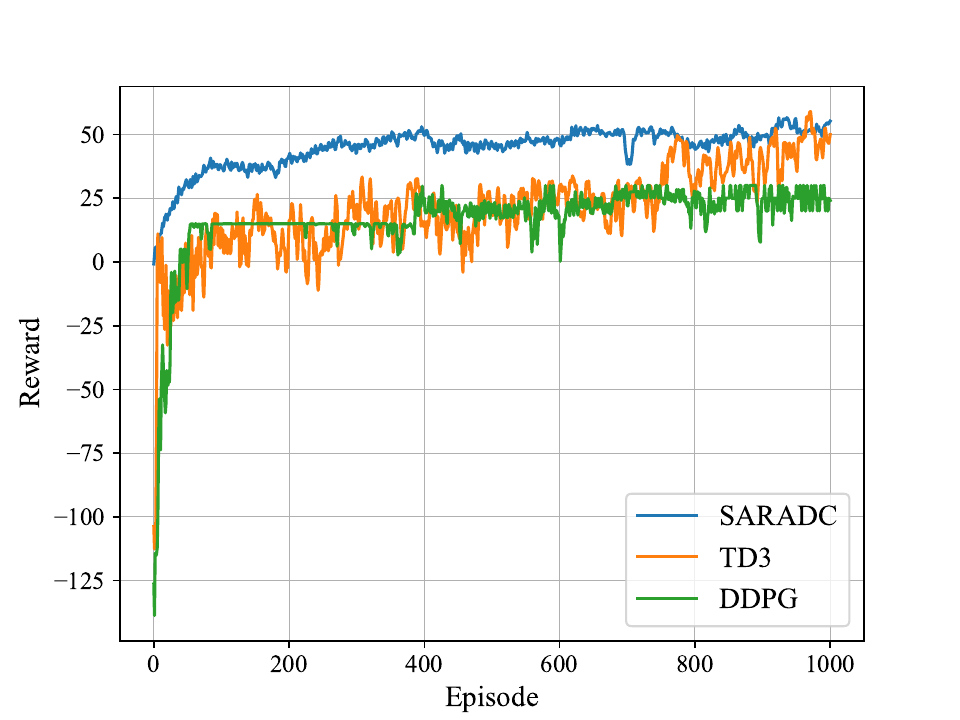}
	\caption{Reward during training process}
	\label{fig3}
\end{figure}
We evaluate SARADC algorithm using Python 3.7. The simulation scenario involves 5 vehicles in a 1000×1000 ${m^2}$ area at a constant speed of 36 $km/h$. 
The channel condition updates every 5ms,
balancing computational complexity and reflecting realistic changes in CSI, providing a stable training environment for the DRL model.
The path loss is 128.1 + 37.6 $\log \left( d \right)$, where $d$ represents the distance between vehicles and BSs. 
Appropriate values for the learning rate $lr$, decay rate $\partial $, and other parameters are set, as shown in Table~\ref{tab1}.

To evaluate the performance of the proposed SARADC algorithm with flexible DC, we compare our algorithm against five baselines as below: a semantic-aware resource allocation (SARA) based on deep deterministic policy gradient (DDPG) RL with flexible DC; a SARA based on twin delayed deep deterministic policy gradient (TD3) RL with flexible DC; Ran\_SC, SARA with random choice and flexible DC; Fixed and Random, SARA based on PPO RL with fixed DC and random DC; DDPG\_NO\_SC, a resource allocation algorithm with unit bit using DDPG RL with flexible DC, excluding semantic symbols.

\begin{figure*}[htbp]
	\centering
	\subfloat[HSSE of vehicles and WiFi]
	{\includegraphics[width=0.31\textwidth]{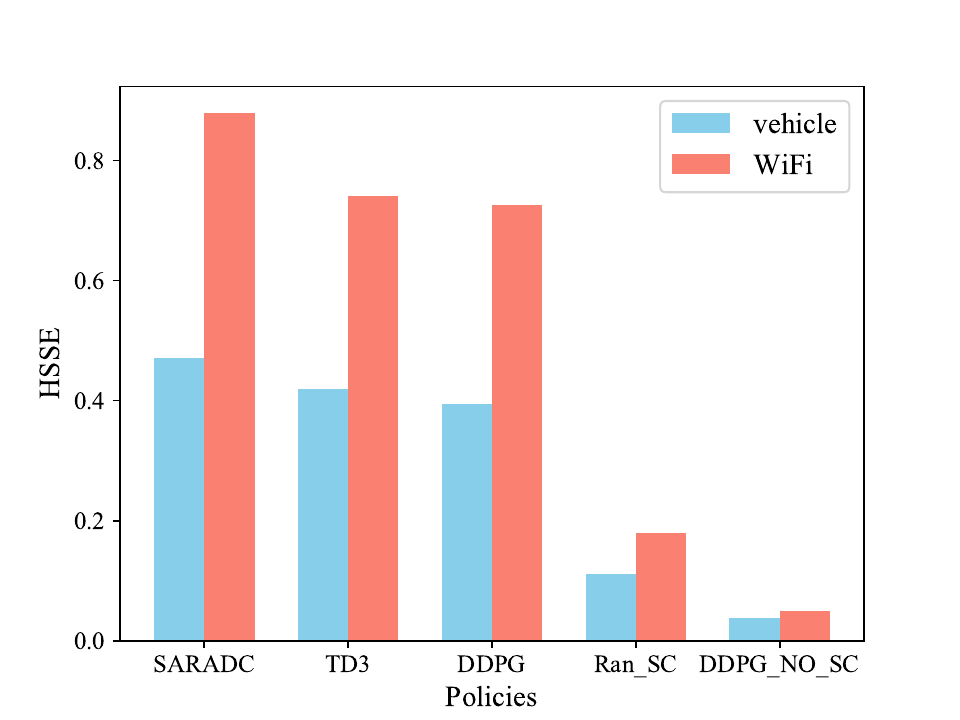}}
	\subfloat[HSSE v.s. transforming factor]
	{\includegraphics[width=0.31\textwidth]{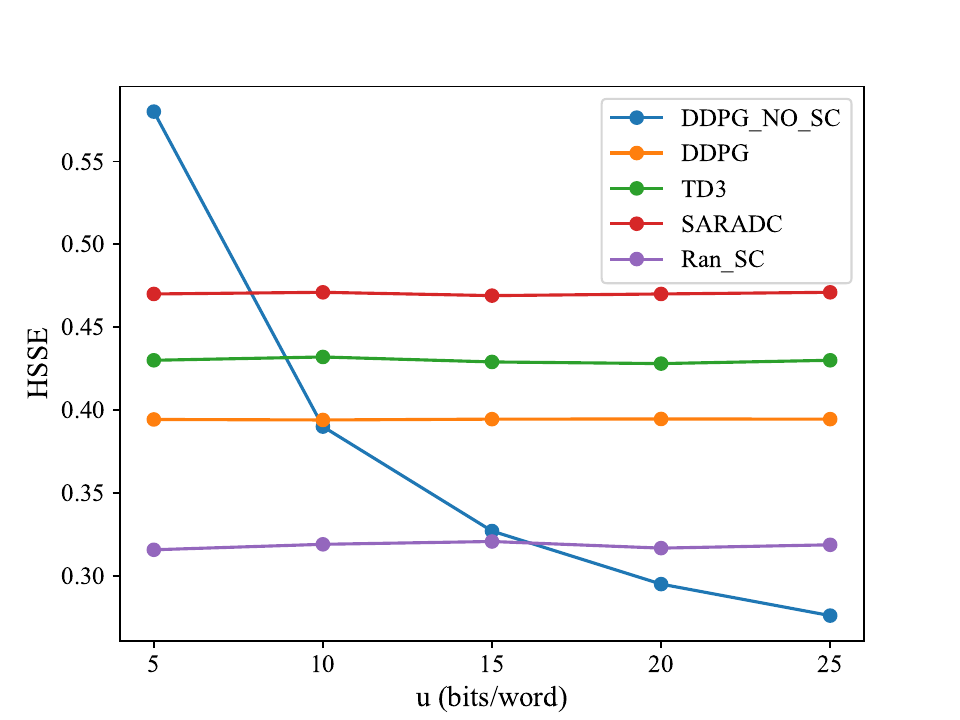}}
	\subfloat[The ST of vehicles and WiFi]
	{\includegraphics[width=0.31\textwidth]{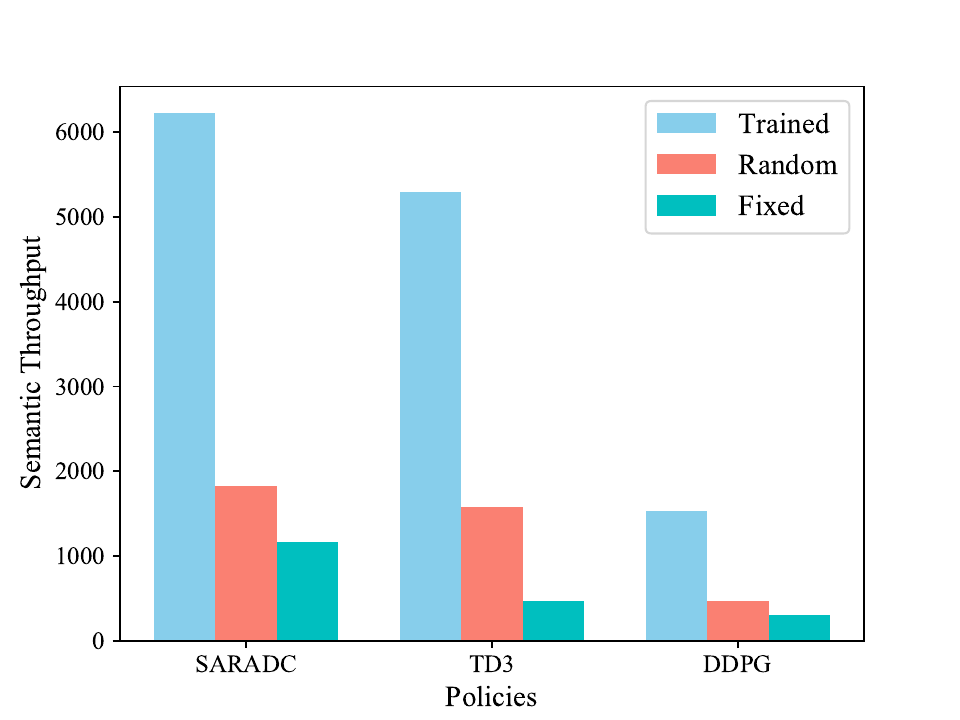}}
	\caption{Variations of HSSE and ST under different algorithms}
	\label{fig4}
\end{figure*}

Figure \ref{fig3} shows the training rewards. The SARADC algorithm converges faster and more stably, yielding superior cumulative rewards. Compared to TD3 and DDPG, SARADC with PPO DRL ensures stability by constraining policy changes and effectively balancing exploration and exploitation, expediting convergence for complex tasks like SARA and DC allocation.

Fig. \ref{fig4}(a) illustrates the HSSE of vehicles and WiFi users under different algorithms. 
WiFi users generally show higher HSSE due to higher bandwidth demands. 
SARADC excels in HSSE, outperforming TD3, DDPG and Ran\_SC due to the stability of PPO's effective exploration and exploitation.
Algorithms incorporating semantic information outperform DDPG\_NO\_SC due to efficient resource utilization of meaningful data. 
Ran\_SC performs better than DDPG\_NO\_SC because it leverages semantic information but is still inferior to SARADC due to the lack of dynamic adaptability.

Fig. \ref{fig4}(b) shows the relationship between HSSE and ${\mu}$.
As ${\mu}$ increases, HSSE remains constant for SARADC and other three semantic-information-based algorithms because transmitting semantic information is independent of ${\mu}$. 
The HSSE of DDPG\_NO\_SC decreases with increasing ${\mu}$. 
When ${\mu}$ is more than 8 bits/word (i.e., a word is encoded by more than 8 bits), our algorithm outperforms traditional algorithms, highlighting the importance of semantic source coding schemes.

Fig. \ref{fig4}(c) shows the total ST of vehicles and WiFi users under different algorithms and DC strategies.
Fixed or random DC strategies lead to lower total ST due to less flexibility and adaptability. Our algorithm consistently outperforms TD3 and DDPG regardless of the DC strategy, as PPO DRL enables faster identification of optimal strategies.
\section{Conclusion}
This letter proposes a SARADC framework tailored for 5G-V2X HetNets using DRL with PPO, integrating semantic communication into high-speed vehicular networking. 
%We compared our algorithm with five baseline methods. SARADC demonstrates excellent performance in terms of HSSE and ST, proving the \textcolor{red}{effectiveness of PPO and semantic communication.} 
The conclusions are summarized as below, firstly, SARADC achieves significantly higher HSSE by leveraging semantic data, ensuring efficient resource utilization. Secondly, fixed or random DC results in lower total system throughput due to their lack of adaptability, whereas our flexible DC adapts to demand, boosting performance. Lastly, using semantic information transmission shows noticeable advantages when data is mapped to more than 8 bits with traditional encoding.

\end{document}